\documentclass[aps,prl,twocolumn,floats,epsfig,showpacs]{revtex4-1}

\usepackage{bbm}
\usepackage{amssymb}
\usepackage{ifpdf}
\usepackage{color}
\usepackage{graphicx,epsfig}

\newcommand{\1}{{\mathbbm{1}}}

\begin{document}

\title{Real-Time Simulation of Large Open Quantum Spin Systems driven by 
Measurements}
\author{D.\ Banerjee$^1$, F.-J.\ Jiang$^2$, M.\ Kon$^3$, and U.-J.\ Wiese$^1$}
\affiliation{$^1$Albert Einstein Center, Institute for Theoretical Physics, 
Bern University, 3012 Bern, Switzerland \\
$^2$ Department of Physics, National Taiwan Normal University
88, Sec.\ 4, Ting-Chou Rd., Taipei 116, Taiwan \\
$^3$ Department of Mathematics, Boston University, Boston, Massachusetts, 
U.S.A.}

\begin{abstract}
We consider a large quantum system with spins $\frac{1}{2}$ whose dynamics 
is driven entirely by measurements of the total spin of spin pairs. This gives
rise to a dissipative coupling to the environment. When one averages over the 
measurement results, the corresponding real-time path integral does not suffer 
from a sign problem. Using an efficient cluster algorithm, we study the 
real-time evolution of a 2-d Heisenberg antiferromagnet, which is driven to a 
disordered phase, either by sporadic measurements or by continuous monitoring 
described by Lindblad evolution.
\end{abstract}

\maketitle

Simulating the real-time evolution of large quantum systems is a notoriously
hard problem. On the one hand, due to the enormous dimension of the Hilbert 
space, which grows exponentially with the system size, diagonalizing the 
Hamiltonian is impossible in practice. On the other hand, the configurations 
contributing to the real-time path integral have complex weights, which 
prevents the application of the Monte Carlo method based on importance sampling.
While this method often works extremely well for Euclidean time simulations of 
quantum systems in thermal equilibrium, it fails for real-time simulations, due 
to a severe sign or complex weight problem. A notable exception are gapped 1-d 
systems with small entanglement, for which the matrix product states underlying 
the density matrix renormalization group \cite{Whi92,Sch05} provide a good basis
for simulating the real-time evolution, at least for moderate time intervals
\cite{Vid03,Whi04,Ver04,Zwo04,Dal04,Bar09,Piz13}. Also Euclidean time 
simulations may suffer from severe sign problems, for example, in fermionic
systems away from half-filling or in the presence of frustrating interactions. 
Some sign problems even fall in the complexity class of NP-complete problems 
\cite{Tro05}, which can be solved in polynomial time on a hypothetical 
``non-deterministic'' computer, but not on an ordinary deterministic computer 
(unless NP would unexpectedly coincide with the complexity class P). This means 
that a general method for solving sign problems is unlikely to exist, and that 
these problems should thus be addressed on a case by case basis. In fact, 
several severe sign problems have been solved completely using the 
meron-cluster algorithm \cite{Bie95,Cha99} or the fermion bag approach 
\cite{Cha10,Cha12,Cha13}.

It is not surprising that classical computers have problems simulating quantum 
systems, in particular, in real time. The entanglement inherent in complex 
quantum phases is not easily representable, let alone computable, as classical 
information. For this reason, as early as 1982 Feynman proposed using 
specifically designed quantum devices to mimic quantum systems that are 
difficult to simulate classically \cite{Fey82}. Since the ground-breaking 
experimental realization of Bose-Einstein condensation \cite{And95,Dav95}, the 
fields of atomic physics and quantum optics have undergone impressive 
development. The degree to which ultracold atomic systems can be engineered and 
controlled is truly remarkable, and Feynman's vision of quantum simulators is
becoming a reality. For example, the bosonic 
Hubbard model has been implemented with exquisitely well-controlled ultracold 
atoms in an optical lattice \cite{Gre02}, and several aspects of this quantum 
simulation have been verified by comparison with accurate quantum Monte Carlo 
simulations \cite{Tro10}. Digital \cite{Llo96} and analog \cite{Jak98} quantum 
simulators are widely discussed in atomic and condensed matter physics 
\cite{Cir12,Lew12,Blo12,Bla12,Asp12,Hou12}, and more recently also in a 
particle physics context \cite{Kap11,Szi11,Zoh12,Ban12,Ban13,Zoh13,
Tag13a,Tag13b,Wie13}.

While quantum simulators are gradually becoming available, they are far from 
being universally applicable, and they are not yet precision instruments. 
Hence, simulating the real-time evolution of large quantum systems on classical
computers remains an important challenge. Since isolated quantum systems tend 
to evolve into complicated entangled states such as those of Schr\"odinger's
cat, it will in general be extremely difficult to compute them classically.
In the real world, Schr\"odinger cat states usually do not arise, because 
quantum systems suffer from decoherence by coupling to their environment, and 
thus behave more classically. It should hence be easier to simulate quantum 
systems in the presence of an environment. Here we develop a method to simulate
the real-time evolution of large quantum spin systems whose dynamics are 
entirely driven by measurements of the total spin $(\vec S_x + \vec S_y)^2$ of 
pairs of spins $\frac{1}{2}$ at adjacent positions $x$ and $y$. The measurements
give rise to a dissipative coupling to the environment, which drives the system 
from an initial state to a new equilibrium. Remarkably, when one averages over 
the measurement results, the sign problem is eliminated and the dynamics can be 
addressed with an efficient cluster algorithm. This is the 
first time that the real-time evolution of a large strongly coupled quantum 
system can be simulated over arbitrarily long time intervals in any spatial 
dimension. The dissipative measurement process that drives the time-evolution 
may even be realizable in optical lattice experiments. The control of quantum 
systems by measurements is investigated in \cite{Sug05,Pec06}, and dynamical 
phenomena in out-of-equilibrium quantum systems are discussed in 
\cite{Cor74,Ber02,Aar02,Ber08,Die08,Dal09,Ver09,Die10,Mue12,Sie13,DeG13}. 
Measurements have also been suggested as a resource for quantum computation 
\cite{Rau01,Nie01,Chi02,Ali04}. In non-relativistic quantum mechanics, the path 
integral representation of measurement processes has been discussed in 
\cite{Men79,Cav86}.

Let us consider a general quantum system with a (possibly time-dependent) 
Hamiltonian, whose real-time evolution from $t_k$ to $t_{k+1}$ is described by 
the time-evolution operator $U(t_{k+1},t_k) = U(t_k,t_{k+1})^\dagger$. At time 
$t_k$ ($k \in \{1,2,\dots,N\}$) we assume an observable $O_k$ is measured and an
eigenvalue $o_k$ is obtained as the measurement result. The Hermitean operator 
$P_{o_k}$ projects on the subspace of the Hilbert space spanned by the 
eigenvectors of $O_k$ with eigenvalue $o_k$. Starting from an initial 
density matrix $\rho_0 = \sum_i p_i |i\rangle\langle i|$ (with 
$0 \leq p_i \leq 1$, $\sum_i p_i = 1$) at time $t_0$, the probability of 
reaching a final state $|f\rangle$ at time $t_f$, after a sequence of $N$ 
measurements with results $o_k$, is then given by \cite{Gri84}
\begin{eqnarray}
\label{probrhof}
&&p_{\rho_0 f}(o_1,o_2,\dots,o_N) = \nonumber \\
&&\sum_i \langle i|U(t_0,t_1) P_{o_1} U(t_1,t_2) P_{o_2} \dots P_{o_N} U(t_N,t_f)
|f\rangle \nonumber \\
&&\langle f|U(t_f,t_N) P_{o_N} \dots P_{o_2} U(t_2,t_1) P_{o_1} U(t_1,t_0)|i\rangle
p_i.
\end{eqnarray}
The matrix elements of both the time-evolution and the projection operators are 
in general complex, thus leading to a severe sign problem in Monte Carlo
simulations. As we have argued above, classical measurements disentangle the 
quantum system, at least to some extent, and should thus alleviate the sign 
problem. For simplicity, we now consider quantum systems whose time-evolution 
is entirely driven by measurements, i.e.\ $U(t_k,t_{k+1}) = \1$. By inserting 
complete sets of states $\sum_{n_k}|n_k\rangle\langle n_k| = \1$ into the first 
factor and independently $\sum_{n_k'}|n_k'\rangle\langle n_k'| = \1$ into the 
second factor in eq.\ (\ref{probrhof}), between the times $t_k$, one arrives at 
a real-time path integral along the Keldysh contour leading from $t_0$ to $t_f$ 
and back \cite{Sch61,Kel65}. In the doubled Hilbert space of states 
$|n_k n_k'\rangle$, encompassing both pieces of the Keldysh contour,
\begin{eqnarray}
&&p_{\rho_0 f}(o_1,o_2,\dots,o_N) = \nonumber \\
&&\sum_i p_i \langle i i|(P_{o_1} \otimes P_{o_1}^*)(P_{o_2} \otimes P_{o_2}^*) \dots
(P_{o_N} \otimes P_{o_N}^*)|f f\rangle = \nonumber
\end{eqnarray} \vspace{-0.8cm}
\begin{equation}
\sum_i p_i \sum_{n_1,n_1'} \dots \!\!\!\!\!\! \sum_{n_{N-1},n_{N-1}'} 
\prod_{k=1}^N \langle n_{k-1} n_{k-1}'|P_{o_k} \otimes P_{o_k}^*|n_k n_k'\rangle.
\end{equation}
We use the notation
$\langle n_{k-1} n_{k-1}'|P_{o_k} \otimes P_{o_k}^*|n_k n_k'\rangle =
\langle n_{k-1}|P_{o_k}|n_k \rangle \langle n_{k-1}'|P_{o_k}|n_k'\rangle^*$,
$\langle n_0 n_0'| = \langle i i|$, and $|n_N n_N'\rangle = |f f\rangle$. 
We also consider the probability $p_{\rho_0 f}$ of reaching the final state 
$|f\rangle$ irrespective of the intermediate measurement results,
\begin{eqnarray}
\hspace{-2.2cm}
p_{\rho_0 f} = \sum_{o_1} \sum_{o_2} \dots \sum_{o_N} p_{\rho_0 f}(o_1,o_2,\dots,o_N) 
\nonumber
\end{eqnarray} \vspace{-0.8cm}
\begin{equation}
\hspace{0.65cm} = \sum_i p_i \sum_{n_1,n_1'} \dots \!\!\!\!\!\! \sum_{n_{N-1},n_{N-1}'} 
\prod_{k=1}^N \langle n_{k-1} n_{k-1}'|\widetilde P_k|n_k n_k'\rangle,
\end{equation}
where $\widetilde P_k = \sum_{o_k} P_{o_k} \otimes P_{o_k}^*$ is obtained by summing
over all possible measurement results $o_k$ at time $t_k$. 

Besides the process of sporadic measurements, let us also consider 
quantum systems that are continuously monitored by their environment. This
situation is characterized by a set of Lindblad operators \cite{Kos72,Lin76} 
$L_{o_k} = \sqrt{\varepsilon \gamma} P_{o_k}$ (related to Kraus operators 
\cite{Kra83}) that obey 
$(1 - \varepsilon \gamma N) \1 + \sum_{k,o_k} L_{o_k}^\dagger L_{o_k} = \1$. Here
$\gamma$ determines the probability of measurements per unit time, and
$k \in \{1,2,\dots,N\}$ labels the operators $O_k$ (with eigenvalues $o_k$) that
can induce quantum jumps at any moment in time. In the continuous time 
limit, $\varepsilon \rightarrow 0$, and in the absence of a Hamiltonian, the 
time-evolution of the density matrix is then determined by the Lindblad equation
\begin{eqnarray}
\partial_t \rho&=&\frac{1}{\varepsilon} \sum_{k,o_k} \left(
L_{o_k} \rho L_{o_k}^\dagger - \frac{1}{2} L_{o_k}^\dagger L_{o_k} \rho - 
\frac{1}{2} \rho  L_{o_k}^\dagger L_{o_k} \right) \nonumber \\
&=&\gamma \sum_k (\sum_{o_k} P_{o_k} \rho P_{o_k} - \rho).
\end{eqnarray}

As a simple example, let us first consider two spins $\frac{1}{2}$, $\vec S_x$
and $\vec S_y$, forming total spin $S$ eigenstates $|S S^3\rangle$ (with
3-component $S^3$): $|1 1\rangle = \uparrow\uparrow$,
$|1 0\rangle = \frac{1}{\sqrt{2}}(\uparrow\downarrow + \downarrow\uparrow)$,
$|1 -1\rangle = \downarrow\downarrow$, and
$|0 0\rangle = \frac{1}{\sqrt{2}}(\uparrow\downarrow - \downarrow\uparrow)$.
The projection operators corresponding to a measurement 1 or 0 of the total spin
are then given by $P_1 = |1 1\rangle \langle 1 1| + |1 0\rangle \langle 1 0| +
|1 -1\rangle \langle 1 -1|$ and $P_0 = |0 0\rangle \langle 0 0|$, such that
\begin{equation}
P_1 = \left(\begin{array}{cccc} 
1 & 0 & 0 & 0 \\ 0 & \frac{1}{2} & \frac{1}{2} & 0 \\
0 & \frac{1}{2} & \frac{1}{2} & 0 \\ 0 & 0 & 0 & 1 \end{array}\right), \quad
P_0 = \left(\begin{array}{cccc} 
0 & 0 & 0 & 0 \\ 0 & \frac{1}{2} & - \frac{1}{2} & 0 \\
0 & - \frac{1}{2} & \frac{1}{2} & 0 \\ 0 & 0 & 0 & 0 \end{array}\right).
\end{equation}
The negative entries in $P_0$ give rise to a sign problem in the corresponding
real-time path integral. We quantize the spins in the 3-direction, with 
$s_x = \pm \frac{1}{2}$ denoting the eigenvalues of $S_x^3$. In the doubled 
Hilbert space of states
$|n_k n_k'\rangle = |s_{x,k} s_{y,k} s'_{x,k} s'_{y,k}\rangle$ one then obtains
\begin{eqnarray}
\label{rules}
&&\langle s_{x,k} s_{y,k} s'_{x,k} s'_{y,k}|\widetilde P|
s_{x,k+1} s_{y,k+1} s'_{x,k+1} s'_{y,k+1}\rangle = \nonumber \\
&&(\delta_{s_{x,k},s_{x,k+1}} \delta_{s_{y,k},s_{y,k+1}}
\delta_{s'_{x,k},s'_{x,k+1}} \delta_{s'_{y,k},s'_{y,k+1}} \nonumber \\
&&+ \delta_{s_{x,k},s_{y,k+1}} \delta_{s_{y,k},s_{x,k+1}}
\delta_{s'_{x,k},s'_{y,k+1}} \delta_{s'_{y,k},s'_{x,k+1}})/2.
\end{eqnarray}
All matrix elements of $\widetilde P = P_1 \otimes P_1^* + P_0 \otimes P_0^*$ are
non-negative. The Kronecker $\delta$-functions encode loop-cluster rules for 
binding parallel spins together \cite{Eve93,Wie94}. The Lindblad process is the
continuous-time limit of the discrete measurement process, which can be 
simulated directly in continuous time \cite{Bea96}. Remarkably, the resulting 
cluster algorithm allows very efficient real-time simulations, without 
encountering a sign problem.

The simple two-spin system is easily extended to a large system in any 
dimension. We investigate a system of quantum spins $\frac{1}{2}$ on a 
square lattice of size $L \times L$ with periodic boundary conditions. To
define an initial density matrix $\rho_0 = \exp(- \beta H)$, we consider the 
antiferromagnetic Heisenberg Hamiltonian,
$H = J \sum_{\langle xy \rangle} \vec S_x \cdot \vec S_y$, which is used only to 
prepare an ensemble of initial states, not to evolve it further in time. The 
real-time evolution is again driven entirely by the measurement of the total 
spin $(\vec S_x + \vec S_y)^2$ of nearest-neighbor spin pairs. In a first step, 
all pairs of neighboring spins separated in the 1-direction, at $x = (x_1,x_2)$ 
and $y = (x_1+1,x_2)$ with even $x_1$, are measured simultaneously. In a second 
step, the pairs at $(x_1,x_2)$ and $(x_1,x_2+1)$ with even $x_2$, which are 
separated in the 2-direction, are examined. In a third and fourth measurement 
step, the total spins of pairs with odd $x_1$ and $x_2$ are being measured. Then
the same four-step measurement process is repeated an arbitrary number of times 
$M$, such that the total number of measurements is $N = 4 M$. This particular 
measurement sequence was chosen arbitrarily and can be replaced by any other 
one. For the Lindblad process the ordering of the measurements is irrelevant.
Together with the Keldysh contour, the Euclidean time interval 
$[0,\beta]$ (where $T = 1/\beta$ is the temperature) forms a closed contour in 
the complex time plane. The clusters, which are closed loops extending through 
both real and Euclidean time, are updated by simultaneously flipping all spins 
belonging to the same cluster with probability $\frac{1}{2}$. Remarkably, as a 
global consequence of eq.\ (\ref{rules}), the clusters, and thus also the spin 
configurations that contribute to the real-time path integral, are identical on 
both parts of the Keldysh contour, i.e.\ $s_{x,k} = s'_{x,k}$. Then
eq.\ (\ref{rules}) simplifies to
\begin{eqnarray}
&&\langle s_{x,k} s_{y,k} s_{x,k} s_{y,k}|\widetilde P|
s_{x,k+1} s_{y,k+1} s_{x,k+1} s_{y,k+1}\rangle = \nonumber \\
&&(\delta_{s_{x,k},s_{x,k+1}} \delta_{s_{y,k},s_{y,k+1}} + 
\delta_{s_{x,k},s_{y,k+1}} \delta_{s_{y,k},s_{x,k+1}})/2 = \nonumber \\
&&\langle s_{x,k} s_{y,k}|P_1|s_{x,k+1} s_{y,k+1}\rangle.
\end{eqnarray}

\begin{figure}[tbp]
\includegraphics[width=0.2\textwidth]{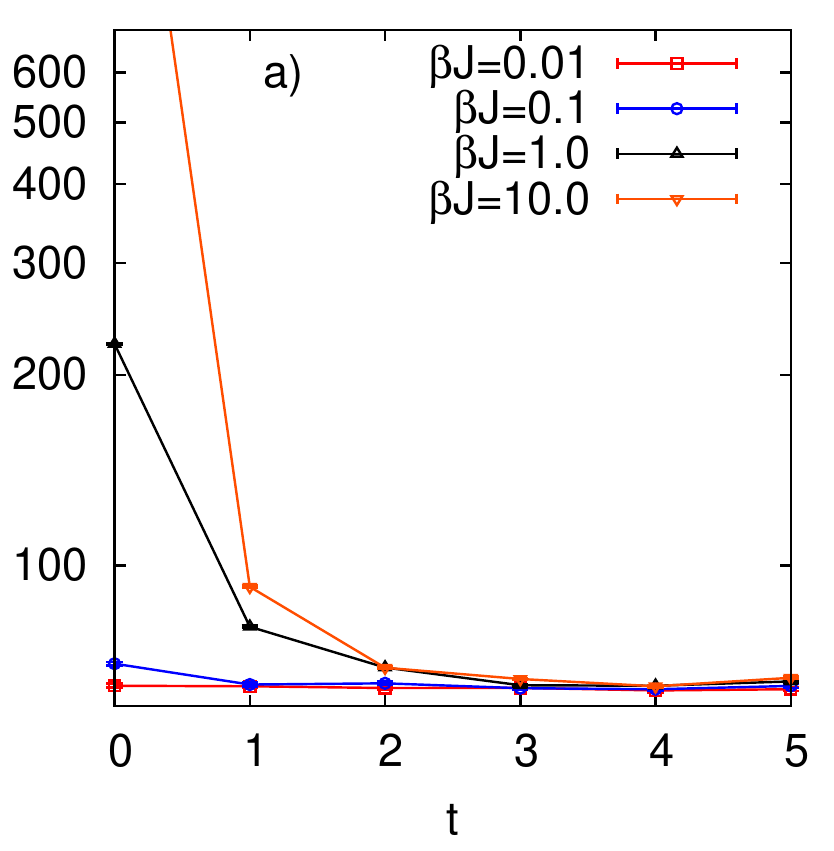}
\includegraphics[width=0.27\textwidth]{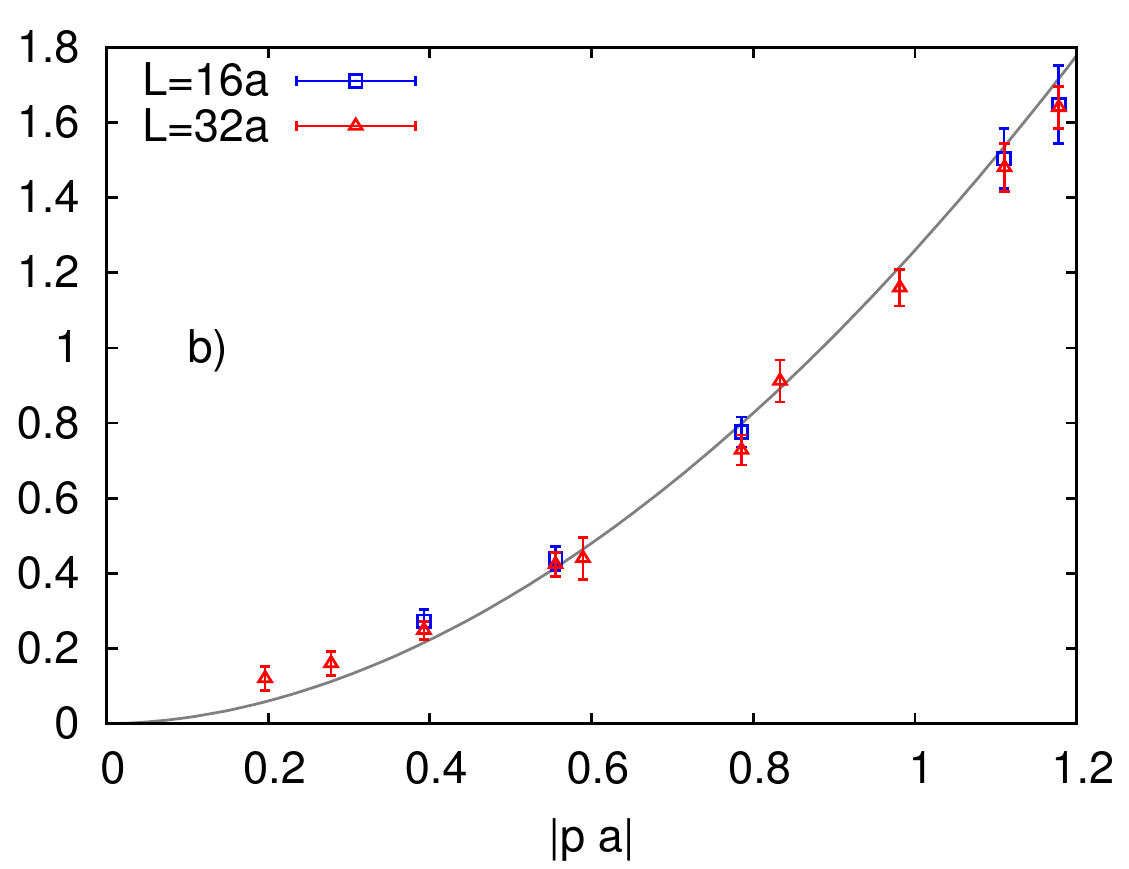} \\
\includegraphics[width=0.47\textwidth]{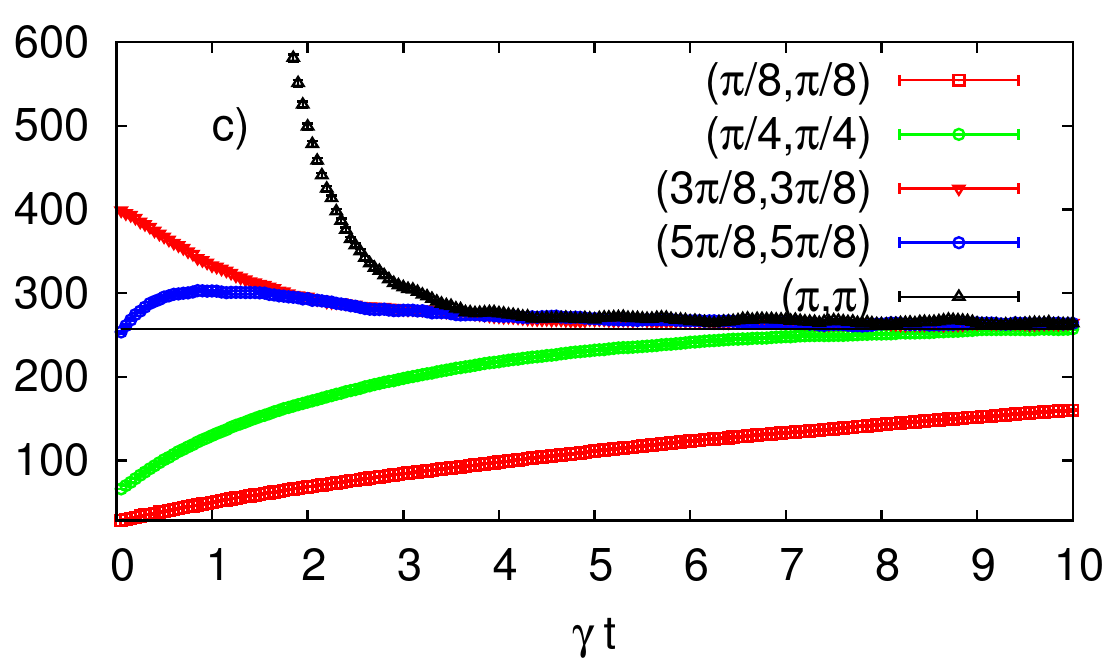}
\caption{[Color online] \textit{a) Real-time evolution of 
$\langle M_s^2 \rangle$ driven by discrete measurements, for 
$\beta J = 0.01$, $0.1$, $1$, and $10$, for $L = 16 a$. b) Inverse equilibration
time $1/[\gamma \tau(p)]$ as a function of $|p|$ for $L = 16 a$, $\beta J = 40$,
and $L = 32 a$, $\beta J = 80$. c) Evolution of the Fourier modes 
$\langle |\widetilde S(p)|^2 \rangle$ for the Heisenberg antiferromagnet driven 
by a continuous Lindblad process.}}
\end{figure}

We have investigated the real-time evolution of initial state ensembles 
corresponding to the 2-d square lattice Heisenberg antiferromagnet. While the 
uniform magnetization, $\vec M = \sum_x \vec S_x$, i.e.\ the total spin, is 
conserved in the measurement process, the staggered magnetization, 
$M_s = \sum_x (-1)^{x_1+x_2} S^3_x$, as well as the other Fourier modes
$\widetilde S(p) = \sum_x S^3_x \exp(i p_1 x_1 + i p_2 x_2)$, $p = (p_1,p_2)$, 
are affected by the measurements. Fig.\ 1a shows the staggered magnetization 
squared, averaged over the ensemble of final states $|f\rangle$ that results 
after $N$ discrete measurements, for systems with different initial 
temperatures. They are quickly driven to a new equilibrium ensemble. 

In order to study the equilibration process in more detail, we now consider 
continuous Lindblad evolution, from an initial ensemble at low temperature 
$\beta J = 5 L/2 a$, where $a$ is the lattice spacing. Fig.\ 1c shows the 
real-time evolution of the Fourier modes
\begin{equation}
\langle |\widetilde S(p)|^2 \rangle \rightarrow A(p) + B(p) \exp(- t/\tau(p)),
\end{equation}
for a variety of momenta $p = (p_1,p_2)$. While the conserved 
magnetization $\vec M$ with momentum $p = (0,0)$ does not equilibrate at all,
low momentum modes equilibrate more slowly than high momentum modes. After a 
short initial phase, the various modes approach the ultimate new equilibrium 
exponentially, with an equilibration time $\tau(p)$. Interestingly, for fixed 
momentum, $\tau(p)$ is almost independent of the spatial volume. Large 
systems equilibrate slowly, because they contain modes of low momentum. For 
small momenta, the equilibration time behaves as 
$1/[\gamma \tau(p)] = C |p a|^r$, $C = 1.26(8)$, $r = 1.9(2)$ (Fig.\ 1b).

Since the measurement process conserves the total spin 
$\vec S = \sum_x \vec S_x$, it does not change the probability distribution 
of the spin associated with the initial density matrix $\rho_0$. The
continuous-time Lindblad process even respects the translation and rotation
symmetries of the lattice. The final density matrix, to which the system is 
driven by the measurements, is constrained by these symmetries, and is
proportional to the unit matrix in each symmetry sector. This finally leads to 
a vanishing correlation length and to $A(p) = L^4/4 (L^2 - 1)$, indicated by 
the horizontal line in Fig.\ 1c. The unit density matrix (restricted to the 
appropriate symmetry sectors) is a stable $T = \infty$ fixed point of any 
Hamiltonian plus Lindbladian dynamics, and thus a universal attractor for the 
ultimate long-term evolution for a large class of dissipative processes.

For the initial density matrix $\rho_0$ of the antiferromagnet,
$\langle M_s^2 \rangle/L^2$ is proportional to $L^2$, indicating spontaneous
symmetry breaking of the $SU(2)$ spin symmetry at zero temperature. By the 
Lindblad process the system is driven to a final density matrix for 
which $\langle M_s^2 \rangle/L^2$ becomes volume-independent, indicating that
the $SU(2)$ spin symmetry is then restored. Consequently, the system must
undergo a phase transition. Since the dissipative Lindblad process
drives the system far out of thermal equilibrium, this phase transition is not
expected to fall to any of the standard dynamical universality classes 
\cite{Hoh77}. Figs.\ 2a,b show $\langle M_s^2 \rangle/L^4$ and the Binder ratio
$\langle M_s^4 \rangle/\langle M_s^2 \rangle^2$ for $\beta J = 2L/3a$. The 
various finite-volume curves for the Binder ratio do not intersect. Instead, 
with increasing volume their inflection point moves to later times. Figs.\ 2c,d 
show the staggered magnetization density ${\cal M}_s$ and the length scale 
$\xi = c/(2 \pi \rho_s)$, where $c$ is the spinwave velocity and $\rho_s$ is 
the spin stiffness, as functions of time, obtained by a fit to
\begin{equation}
\langle M_s(t)^2 \rangle = \frac{{\cal M}_s(t)^2 L^4}{3} 
\sum_{n=0}^3 c_n \left(\frac{\xi(t)}{L}\right)^n,
\end{equation}
which implicitly defines ${\cal M}_s(t)$ and $\xi(t)$. Here the constants 
$c_0 = 1$, $c_1 = 5.7503(6)$, $c_2 = 16.31(2)$, $c_3 = - 84.8(2)$ (which are 
accurately determined at $t = 0$) are assumed to be time-independent. The order 
parameter ${\cal M}_s(t) = {\cal M}_s(0) \exp(-t/\tau)$ (with
${\cal M}_s(0) = 0.30743(1)/a^2$ \cite{San08,Ger09}) decays exponentially with 
$\gamma \tau = 0.240(2)$, which suggests that the phase transition is completed 
only after an infinite amount of time. The length scale $\xi(t)$ (with 
$\xi(0) = 1.459(3)a$ \cite{Ger09}) increases with time, which can be attributed 
to a decrease of $\rho_s$.
\begin{figure}[tbp]
\includegraphics[width=0.239\textwidth]{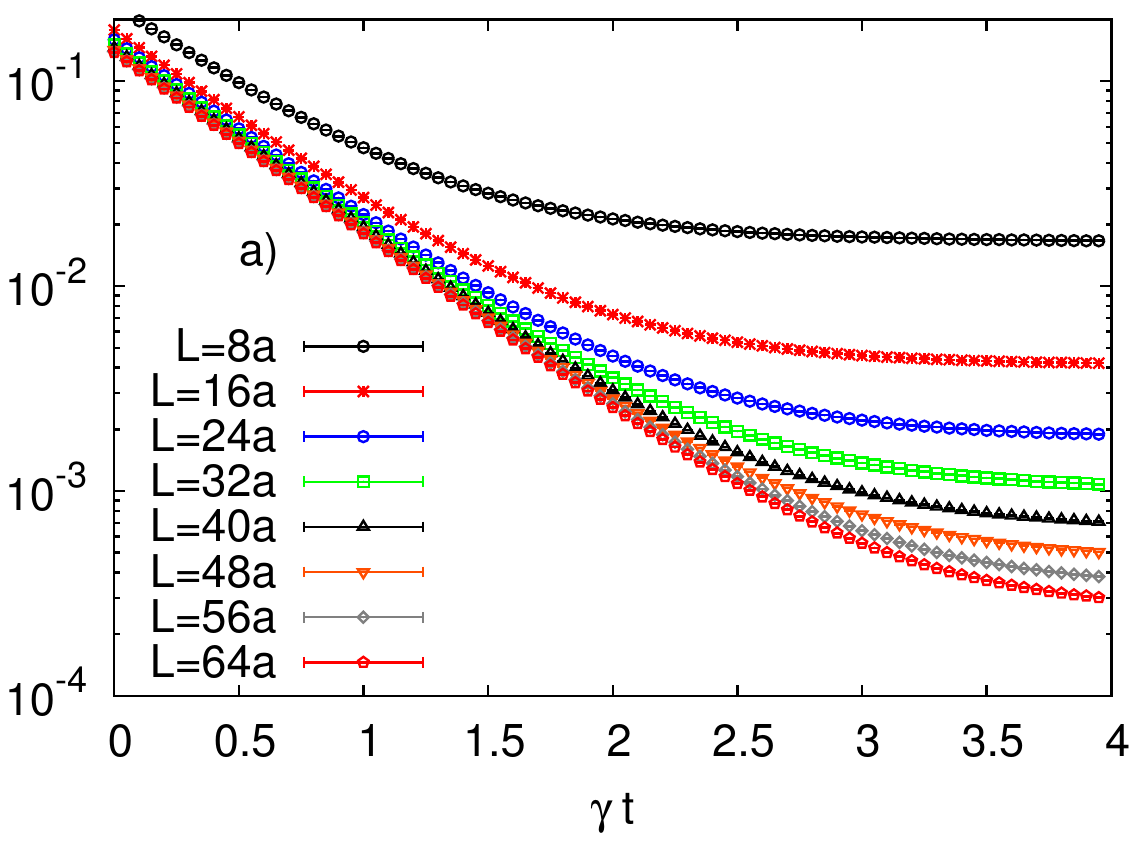}
\includegraphics[width=0.231\textwidth]{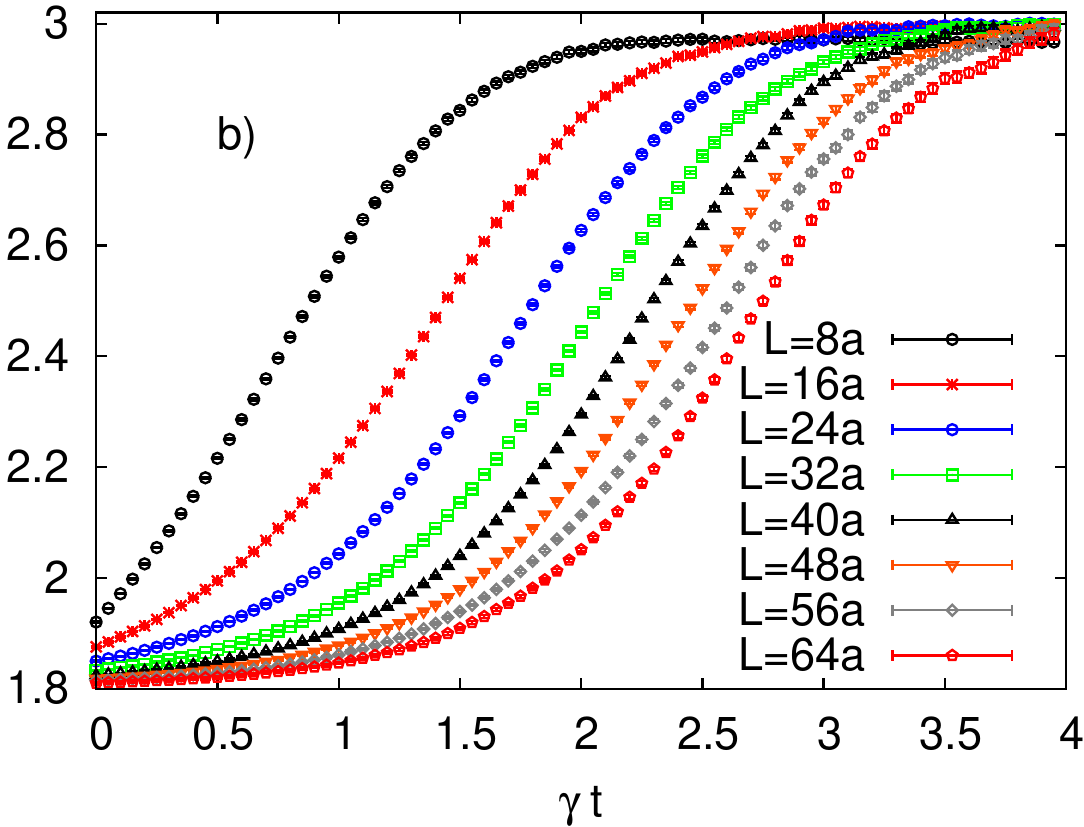} \\
\includegraphics[width=0.235\textwidth]{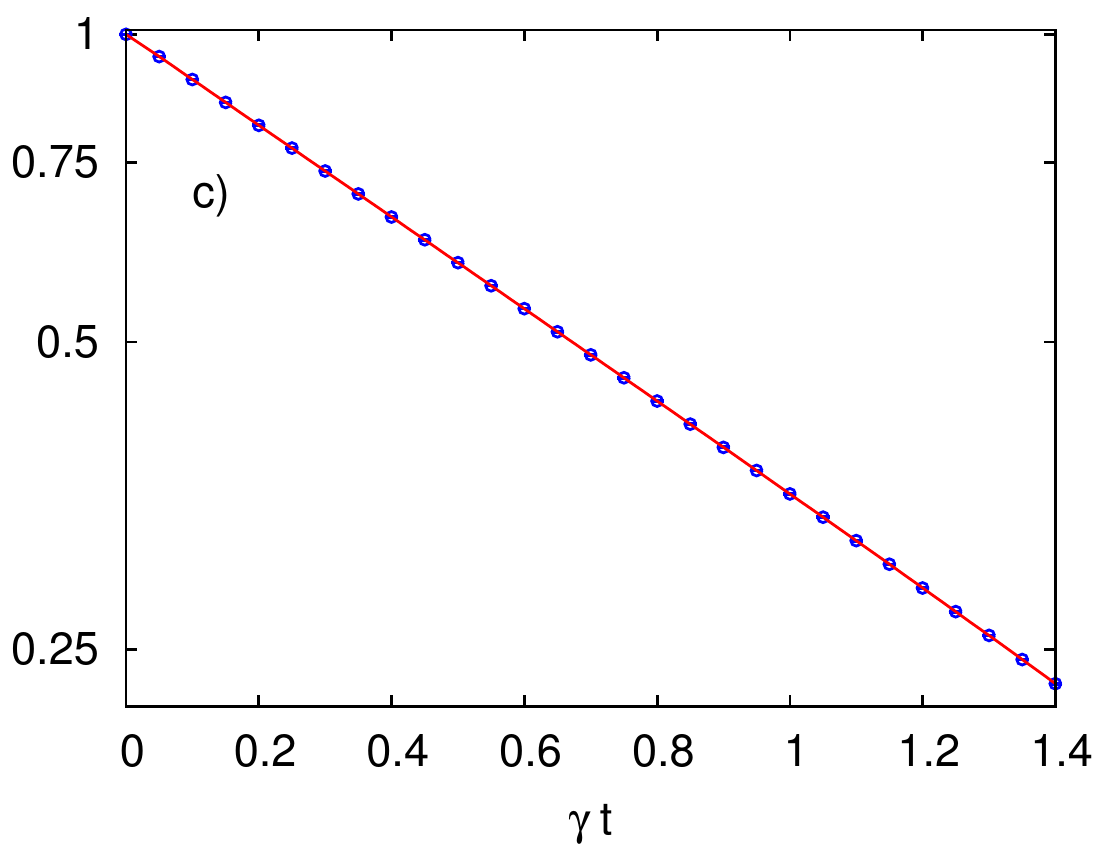}
\includegraphics[width=0.235\textwidth]{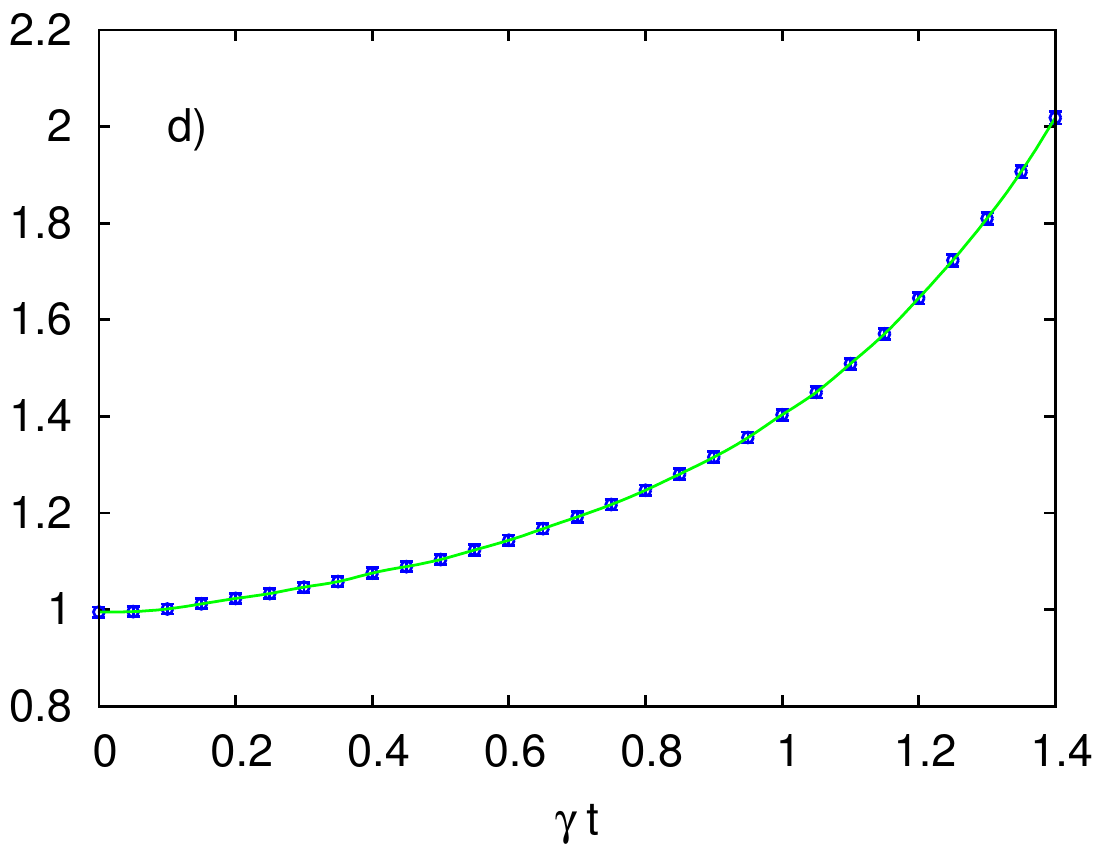}
\caption{[Color online] \textit{a) $\langle M_s^2 \rangle/L^4$ and b) Binder
ratio $\langle M_s^4 \rangle/\langle M_s^2 \rangle^2$ as functions of time for
$L/a = 12,\dots,48$, $\beta J = 8,\dots,30$. Evolution of c) 
${\cal M}_s(t)/{\cal M}_s(0)$ and d) $\xi(t)/\xi(0)$.}}
\end{figure}

By averaging over all measurement results, we have eliminated the sign problem.
When one distinguishes individual measurement results, one encounters a sign
problem. When one measures spin $S = 1$, one obtains
\begin{eqnarray}
\label{breakups}
&&\langle s_{x,k} s_{y,k} s'_{x,k} s'_{y,k}|P_1 \otimes P_1^*|
s_{x,k+1} s_{y,k+1} s'_{x,k+1} s'_{y,k+1}\rangle = \nonumber \\
&&(\delta_{s_{x,k},s_{x,k+1}} \delta_{s_{y,k},s_{y,k+1}}
\delta_{s'_{x,k},s'_{x,k+1}} \delta_{s'_{y,k},s'_{y,k+1}} \nonumber \\
&&+ \delta_{s_{x,k},s_{y,k+1}} \delta_{s_{y,k},s_{x,k+1}}
\delta_{s'_{x,k},s'_{x,k+1}} \delta_{s'_{y,k},s'_{y,k+1}} \nonumber \\
&&+ \delta_{s_{x,k},s_{x,k+1}} \delta_{s_{y,k},s_{y,k+1}}
\delta_{s'_{x,k},s'_{y,k+1}} \delta_{s'_{y,k},s'_{x,k+1}} \nonumber \\
&&+ \delta_{s_{x,k},s_{y,k+1}} \delta_{s_{y,k},s_{x,k+1}}
\delta_{s'_{x,k},s'_{y,k+1}} \delta_{s'_{y,k},s'_{x,k+1}})/4,
\end{eqnarray}
which is always non-negative. Again, the Kronecker $\delta$-functions encode
rules for forming clusters of parallel spins. The four contributions to the 
right-hand side of eq.\ (\ref{breakups}) correspond to four different cluster 
break-ups of the eight contributing spins. When we measure the total spin 
$S = 0$, we obtain
\begin{eqnarray}
&&\hspace{-0.3cm}\langle s_{x,k} s_{y,k} s'_{x,k} s'_{y,k}|P_0 \otimes P_0^*|
s_{x,k+1} s_{y,k+1} s'_{x,k+1} s'_{y,k+1}\rangle = \nonumber \\
&&\hspace{-0.3cm}(s_{x,k}\!-\!s_{y,k})(s_{x,k\!+\!1}\!-\!s_{y,k\!+\!1})
(s'_{x,k}\!-\!s'_{y,k})(s'_{x,k\!+\!1}\!-\!s'_{y,k\!+\!1})/4 \nonumber \\
&&\hspace{-0.3cm}\times \delta_{s_{x,k},-s_{y,k}} \delta_{s_{x,k+1},-s_{y,k+1}}
\delta_{s'_{x,k},-s'_{y,k}} \delta_{s'_{x,k+1},-s'_{y,k+1}},
\end{eqnarray}
which may indeed be negative. In this case, the
Kronecker $\delta$-functions assign anti-parallel spins on neighboring spatial
sites to the same cluster. Interestingly, the resulting sign problem is similar 
to the one that arises for geometrically frustrated quantum antiferromagnets in 
Euclidean time, which has been addressed with a nested cluster algorithm in 
\cite{Jia08}. While this algorithm reduces the sign problem by a factor that is
exponential in the space-time volume, in general it does not solve the problem 
completely. If we distinguish only a few measurement results, and average over 
the other ones, the sign problem remains manageable, and can be solved by the 
nested cluster algorithm.

We have considered quantum spin systems whose real-time evolution is entirely
driven by measurements of the total spin of spin pairs. Remarkably, when one 
averages over the measurement results, the corresponding real-time path integral
is unaffected by the sign problem and has been simulated with a very efficient 
loop-cluster algorithm. Subsequent measurements at discrete times as well as a 
related dissipative continuous-time Lindblad process destroy long-range 
antiferromagnetic correlations of the initial density matrix, and drive the 
system to a new equilibrium with only short-range correlations. Our method can 
be applied to other initial density matrices and can be extended to other 
measurement processes that drive the real-time evolution. It will be interesting
to investigate the real-time evolution of a variety of initial states by 
various dissipative measurement processes, which may even by realizable in
optical lattice experiments with ultracold atoms. A challenging next step will 
be to combine measurements with the real-time evolution driven by a non-trivial 
Hamiltonian.

We like to thank J.\ Berges, S.\ Chandrasekharan, F.\ Niedermayer and P.\ Zoller
for illuminating discussions. MK and UJW thank the CTP at MIT, where this work 
was initiated, for hospitality during a sabbatical. UJW is grateful to J.\ 
Fr\"ohlich for stimulating discussions a long time ago. The research leading to
these results has received funding from the Schweizerischer Na\-tio\-nal\-fonds 
and from the European Research Council under the European Union's Seventh 
Framework Programme (FP7/2007-2013)/ ERC grant agreement 339220.

\end{document}